\documentclass[12pt]{article}
\usepackage{epsfig}
\usepackage{amssymb}
\usepackage{amsmath,amsfonts,amssymb,graphicx}
\newcommand {\be}{\begin{equation}}
\newcommand {\ee}{\end {equation}}
\newcommand{\beq}{\begin{eqnarray}}
\newcommand{\eeq}{\end{eqnarray}}

\begin{document}
\vspace{-2cm}

\title{Neutrino Oscillation in Matter and Parameters $s_{13},\delta_{CP}$}
\author{Leonard S. Kisslinger$^{a}$, Ernest M. Henley$^{b,}$, and
 Mikkel B. Johnson$^{c}$\\
 $^a$Department of Physics, Carnegie Mellon University, Pittsburgh, 
PA 15213 \\
 $^b$Department of Physics, University of Washington, Seattle,
WA 98195 \\
 $^c$Los Alamos National Laboratory, Los Alamos, NM 87545}
\maketitle
\noindent
PACS Indices:11.30.Er,14.60.Lm,13.15.+g\vspace{0.25 in}
\begin{abstract}

We estimate the dependence of $\nu_{\mu}$ to $\nu_{e}$ conversion on
parameters $\theta_{13}$ and $\delta_{CP}$ for several experimental 
facilities studying neutrino oscillations. We use the S-Matrix theory
to estimate $\bar{\nu_e}$ disappearance and compare estimates based on
an older theory being used to extract $\theta_{13}$ from the Double Chooz, 
Daya Bay, and RENO data, to assist in extracting an accurate value for 
$\theta_{13}$ from these projects.  We use values of $\theta_{13}$ within 
known limits, and estimate the dependence of $\nu_{\mu}$ - $\nu_{e}$ CP 
violation (CPV) probability on $\delta_{CP}$ in order to suggest new 
experiments to measure CPV for neutrinos moving in matter.
\end{abstract}
\vspace {0.25 in}

\section{Introduction}

  In our present work we study $\nu_\mu$ to $\nu_e$ neutrino conversion,
$\bar{\nu}_e$ disappearance, and CP violation (CPV) measurements using 
the S-Matrix method for neutrino oscillations. 
The study of CP violation is essential for understanding weak 
interactions.  Almost half a century ago CP violation in weak interactions 
was found in the decay of $K^0_L$ into $\pi^+ + \pi^{-} $ \cite{ccft64} and 
$2 \pi^0$ \cite{ckrw67}, with branching ratios of the order of .001.
The decay $K^0_L \rightarrow \pi^0 + \nu + \bar{\nu}$ is almost entirely
CP violating \cite{ll89} but requires accurate determination of the 
CKM matrix \cite{ckm63} and accurate measurements. See Ref\cite{bu08} for
a review of this experiment and references. There have many other studies 
of CP asymmetries in weak decays: see Ref\cite{blnp11} for a recent study
of $\bar{B}$ radiative decay with references to earlier work on CP violation
in various weak decays.

   In recent years there have been a number of 
experimental studies of neutrino oscillations using neutrino beams from 
accelerators and reactors, and important objectives of these 
experiments are the measurements of $\nu_\mu$ to $\nu_e$ conversion and CPV.
The first study in our present work is an estimate the $\nu_\mu$ to $\nu_e$ 
conversion probability using parameters for the baseline and energy 
corresponding to MiniBooNE\cite{mini}, JHF-Kamioka \cite{jhf}, 
MINOS\cite{minos}, and CHOOZ\cite{chooz}-Double Chooz\cite{dc06}, which are 
on-going projects, although the CHOOZ project does not have a beam of 
$\nu_\mu$ neutrinos.

There have been many recent studies of CP and T symmetries via neutrino 
oscillations for future facilities, e.g., see Refs\cite{dlm11,gms11}, which 
also give references to earlier publications, and the ISS report\cite{ISS} 
on future neutrino facilities. The two main parameters of interest in the 
present work are $\delta_{CP}$, which is essentially unknown, and the
angle $\theta_{13}$. One possible future facility for studying CPV and 
the $\delta_{CP}$ parameter is the LBNE Project, where neutrino beams 
produced at Fermilab would have a baseline of L $\simeq$ 1200 km, being 
detected with deep underground detectors\cite{LBNE,LBNE07}. With the methods 
used in the present work, described below, predictions of CPV with the 
baseline and energies of the LBNE Project have recently been made for 
$\delta_{CP}$ from 90 to 0 degrees\cite{k11}. 

  Although the angle $\theta_{13}$ is not well known, there are limits on 
its value. Recently the T2K collaboration \cite{T2K11} published limits,
and found a best-fit value of $\theta_{13} \simeq $11 degrees, or 
sin$\theta_{13}$ $\equiv s_{13} \simeq$ 0.19, which is one of the values we
use in the present work on neutrino transition probability in Sec. 2.

  The angle $\theta_{13}$ will be measured by the 
Daya Bay experiment\cite{DayaBay,DayaBay11} in China, the Double Chooz 
project\cite{dc11,dcnov11} in France, and RENO\cite{RENO} in Korea,
via $\bar{\nu}_e$ disappearance. A very recent result from the Daya Bay 
project\cite{DB3-7-12} concludes that $s_{13} \simeq 0.15$, and
in our study (Sec. 3) of $\bar{\nu}_e$ disappearance we use the the Daya Bay 
parameters to test the theory. Our results will also be useful for the 
Double Chooz and RENO projects.

Finally, in Sec 4, using the expected range of values for $\theta_{13}$, 
we estimate CPV for $\mu-e$ neutrino oscillation for the entire range of 
$\delta_{CP}$ to help in the planning for future CPV experiments.

A major complication for the determination of
T, CP, and CPT violation is the interaction of neutrinos with matter as
they travel along the baseline. These matter effects have been studied by
a number of theorists. See, e.g., Refs\cite{as97,bgg97,kty02}. One main
objective of the present research is to estimate matter effects
for the MiniBooNE, JHF-Kamioka, MINOS, and CHOOZ facilities, to help find
the values of $\theta_{13}$ and $\delta_{CP}$. 

For the basic interactions, which are CPT invariant for local theories, CP 
and T violation have the same magnitude. With matter effects T and CP are not
directly related. Our present research is an extension of 
our recent work on T reversal violation\cite{hjk11}. In that study we used
the formalism of Ref\cite{ahlo01} for $\nu_e \leftrightarrow \nu_\mu$ TRV,
and that of Ref\cite{f01} for $\nu_e \rightarrow \nu_\mu$ conversion 
probability to calculate the effects of neutrinos moving through matter. 
In the present work we use the notation and formalism of 
Jacobson and Ohlsson\cite{jo04}, who studied possible matter effects for 
CPT violation.

  CP violation in the $a-b$ sector is given by the transition probability,
denoted by $\mathcal{P}(\nu_a \rightarrow \nu_b)$, for a neutrino of flavor
$a$ to convert to a neutrino of flavor $b$; and similarly for antineutrinos
$\bar{\nu}_a,\bar{\nu}_b$.
   The CPV probability differences are defined as
\beq
   \Delta\mathcal{P}^{CP}_{ab}&=& \mathcal{P}(\nu_a \rightarrow \nu_b)
-\mathcal{P}(\bar{\nu}_a \rightarrow \bar{\nu}_b) \; .
\eeq

In our present work we study $\mathcal{P}(\nu_\mu \rightarrow \nu_e)$
and $\mathcal{P}(\nu_\mu \rightarrow \nu_e) -\mathcal{P}({\bar\nu}_\mu 
\rightarrow \bar{\nu}_e)$ since the neutrino 
beams at MiniBooNE, JHF-Kamioka, and MINOS, are muon or anti-muon neutrinos.
We then calculate $\bar{\nu}_e$ disappearance probability for Daya Bay
baseline and energy, comparing our S-Matrix theory to the formula used by 
Daya Bay, Double Chooz and RENO\cite{DayaBay,DayaBay11,dc11,dcnov11,RENO} . 

\section{Transition Probability $ \mathcal{P}(\nu_\mu \rightarrow  
\nu_e)$ and $\bar{\nu_e}$ Disappearance }

In this section we review $\nu_\mu$ to $\nu_e$ oscillation probability
derived from standard S-martix theory and then compare the probability of  
$\bar{\nu_e}$ disappearance derived from this theory to that used by the 
Double Chooz experimental project\cite{dc11,dcnov11}.

\subsection{$\mathcal{P}(\nu_\mu \rightarrow \nu_e)$ Derived Using
S-Matrix Theory}

In this subsection we review the derivation of the probability of a muon 
neutrino to convert to an electron neutrino, $ \mathcal{P}(\nu_\mu 
\rightarrow\nu_e)$, using the notation of Ref\cite{jo04}. We then make an 
estimate of the transition probabilities for
sample accelerator and reactor experiments. Although at the present time no
experiments for CPV are possible, this can serve as a basis for future
experiments. In the next section we give somewhat more accurate calculations
for CPV for the same set of experimental facilities.

As in Refs\cite{ahlo01,jo04} we use the time evolution matrix, $S(t,t_0)$ to 
derive the transition probabilities. For neutrino oscillations the initial
neutrino beam is emitted at time $t_0$, usually taken as $t_0 = 0$, and the
neutrino or converted neutrino is detected at baseline length = $L$ at
time=$t$. Since the neutrinos move with a velocity near that of the speed
of light, at the end of our derivation we take $t-t_0 \rightarrow L$, with
the units c=1.

Given the Hamiltonian, H(t), for neutrinos, the neutrino state at time = $t$
is obtained from the state at time = $t_0$ from the matrix, $S(t,t_0)$, by
\beq
             |\nu(t)> &=& S(t,t_0)|\nu(t_0)> \\
             i\frac{d}{dt}S(t,t_0) &=& H(t) S(t,t_0) \; .
\eeq

Neutrinos (and antineutrinos) are produced as $\nu_a$, where $a$ is the flavor,
$a = e,\; \mu,\; \tau$.  However, neutrinos of definite masses
are $\nu_\alpha$, with $\alpha=1,2,3$. The two forms are connected by a 3 by 3
unitary transformation matrix, $U$: $\nu_a = U \nu_\alpha$, 
where $\nu_a,\nu_\alpha$ are 3x1 column vctors and $U$ is given by 
($sin\theta_{ij} \equiv s_{ij}$)

\beq
 U=\left( \begin{array}{lcr} c_{12}c_{13} &s_{12}c_{13} & s_{13}
e^{-i \delta_{CP}} \\
     -s_{12}c_{23}-c_{12}s_{23}s_{13}e^{i\delta_{CP}} & c_{12}c_{23}-s_{12}
s_{23}s_{13}e^{i\delta_{CP}} & s_{23}c_{13} \\ 
s_{12}s_{23}-c_{12}c_{23}s_{13}e^{i\delta_{CP}} & -c_{12}s_{23}-s_{12}c_{23}
s_{13}e^{i\delta_{CP}} & c_{23}c_{13} \end{array} \right) \nonumber \; ,
\eeq
similar to the CKM matrix for quarks. We use the best fit value\cite{dlm11}
 $s_{23}=0.707$. $\theta_{13}$ and $s_{12} =0.56$, $c_{12} =0.83$ . We use
$s_{13}=0.19$ and $s_{13}$=0.095, as discussed above,
to determine the dependence of $\nu_\mu\rightarrow \nu_e$ conversion,
and CPV on this parameter. We calculate the dependence of 
$\mathcal{P}(\nu_\mu \rightarrow\nu_e)$ 
and $\Delta\mathcal{P}^{CP}_{\mu e}$  on $\delta_{CP}$, as discussed below.
In the vacuum the $S(t,t_0)$ is obtained from
\beq
        S_{ab}(t,t_0)&=& \sum_{j=1}^{3} U_{aj} exp^{i E_j (t-t_0)} U^*_{bj} \; .
\eeq

  Since neutrino beams in neutrino oscillation experiments travel through
matter, and the main neutrino-matter is scattering from electrons, we must
include potential, $V = \sqrt{2} G_F n_e$, for  neutrino electron scattering 
in the earth:
where $G_F$ is the universal weak interaction Fermi constant, and $n_e$ is 
the density of electrons in matter. Using the matter density $\rho$=3 gm/cc, 
the neutrino-matter potential is $V=1.13 \times 10^{-13}$ eV.

  The transition probability $ \mathcal{P}(\nu_\mu \rightarrow\nu_e)$
is obtained from $S_{12}$:
\beq
\label{PueS12}
 \mathcal{P}(\nu_\mu \rightarrow\nu_e) &=& |S_{12}|^2=Re[S_{12}]^2+
Im[S_{12}]^2 \; ,
\eeq
with
\newpage

\beq
\label{S12}
     S_{12} &=& c_{23} \beta -is_{23} a e^{-i\delta_{CP}} A \\
     a &=&  s_{13}(\Delta -s_{12}^2 \delta) \\
     \delta &=& \delta m_{12}^2/(2 E) \\
      \Delta &=&  \delta m_{13}^2/(2 E) \\
     A & \simeq & f(t) I_\alpha* \\ 
       I_\alpha* &=& \int_0^t dt' \alpha^*(t')f(t') \\
     \alpha(t) &=& cos\omega t -i cos 2\theta sin \omega t \\
          f(t) &=& e^{-i \bar{\Delta} t} \\
 2 \omega &=& \sqrt{\delta^2 + V^2 -2 \delta V cos(2 \theta_{12})} \\
         \beta &=& -i sin2\theta sin\omega L \\
         \bar{\Delta} &=& \Delta-(V+\delta)/2 \\
            sin 2\theta&=& s_{12} c_{12} \frac{\delta}{\omega}  \; ,
\eeq
where the neutrino mass differences are $\delta m_{12}^2=7.6 
\times 10^{-5}(eV)^2$ and $\delta m_{13}^2 = 2.4\times 10^{-3} (eV)^2$. 
Note that $\delta \ll \Delta$, and $t\rightarrow L$ for $v_\nu\simeq c$.
From Eqs.(5 to 16):
\beq
\label{Pue}
         Re[S_{12}] &=& s_{23}a[cos(\bar{\Delta}L+\delta_{CP})Im[I_{\alpha*}]
-sin(\bar{\Delta}L+\delta_{CP}) Re[I_{\alpha*}] \nonumber \\
         Im[S_{12}] &=& -c_{23}sin2\theta sin\omega L -s_{23} a
[cos(\bar{\Delta}L+\delta_{CP}) Re[I_{\alpha*}] \nonumber \\
        &&+sin(\bar{\Delta}L+\delta_{CP}) Im[I_{\alpha*}]] 
\eeq

  Using $\delta,\omega \ll \Delta$ one can show that
\beq
\label{I*}
    Re[I_{\alpha*}] &\simeq& \frac{sin \bar{\Delta} L}{\bar{\Delta}} \nonumber \\
    Im[I_{\alpha*}] &\simeq& \frac{1-cos \bar{\Delta} L}{\bar{\Delta}} \; .
\eeq

From Eqs(\ref{Pue},\ref{I*}) we find  
\beq
\label{CPue}
    \mathcal{P}(\nu_\mu \rightarrow\nu_e) &\simeq& (c_{23}s_{12}c_{12}
(\delta/\omega) sin\omega L)^2 +2(s_{23} s_{13})^2(1- cos\bar{\Delta} L) \\
  && +2s_{13}s_{12}c_{12}s_{23}c_{23} (\delta/\omega) sin\omega L
\nonumber \\
  && (cos(\bar{\Delta}L+\delta_{CP})sin\bar{\Delta}L+
sin(\bar{\Delta}L+\delta_{CP})(1-cos\bar{\Delta}L)) \nonumber \;.
\eeq

We use $s_{13}$=.19 and .095 to show the effect of $s_{13}$.
\clearpage

\begin{figure}[ht]
\begin{center}
\epsfig{file=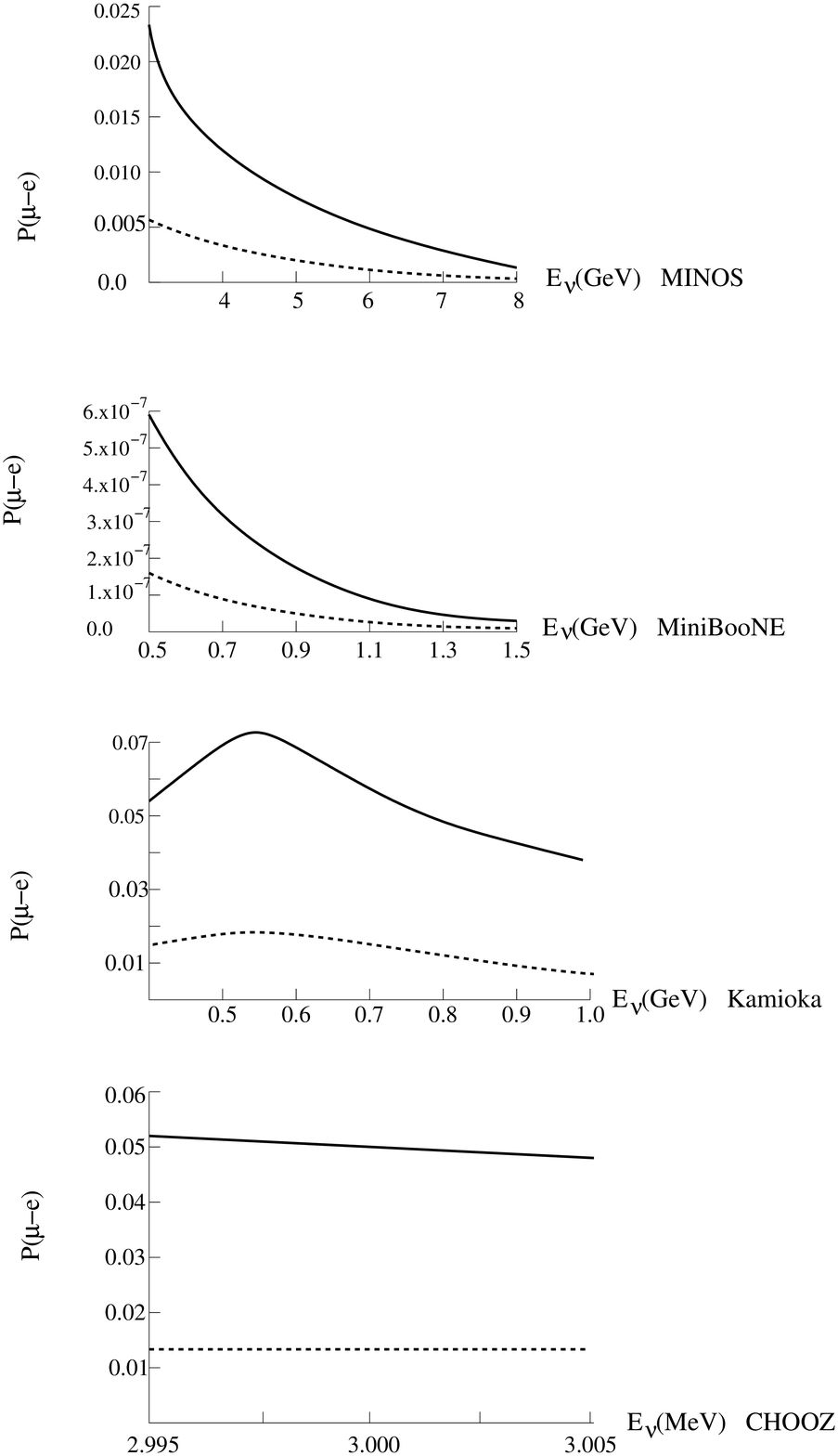,height=18cm,width=12cm}
\end{center}
\caption{\hspace{5mm} The ordinate is $\mathcal{P}(\nu_\mu \rightarrow\nu_e)$ 
for MINOS(L=735 km),
 MiniBooNE(L=500m), JHF-Kamioka(L=295 km), and 
CHOOZ(L=1.03 km).
\hspace {5mm} Solid curve for $s_{13}$=.19 and dashed curve for 
$s_{13}$=.095. The curves are almost independent of $\delta_{CP}$.}
\end{figure}

\clearpage

  From Eq(\ref{CPue}) we obtain the results for $\mathcal{P}(\nu_\mu 
\rightarrow\nu_e)$ shown in Fig.1.
 These results can provide guidance for future 
experiments on CPV via $\nu_\mu \leftrightarrow \nu_e$ oscillation. Note that
in Ref\cite{jhf} $\mathcal{P}(\nu_\mu \rightarrow\nu_e)$ was calculated for
the 295 km JHF-Kamioka project for E=0-2 GeV, and our calculation based on
the theory developed in Refs.\cite{jo04,ahlo01}, finds
 $\mathcal{P}(\nu_\mu \rightarrow\nu_e)$ is in agreement for 
E=.4-1.0 GeV with this earlier estimate.
We calculated $\mathcal{P}(\nu_\mu \rightarrow\nu_e)$ for $\delta_{CP}$
from -$\pi$/2 to $\pi$/2, and the results are almost independent of 
$\delta_{CP}$. The results for CHOOZ are shown in preparing for the following
subsection on $\bar{\nu_e}$ disappearance, even though Double Chooz, Daya
Bay, and RENO projects have beams of $\bar{\nu}_e$ rather than $\nu_\mu$ 
neutrinos. 

\section{$\bar{\nu_e}$ Disappearance Derived Using S-Matrix 
Theory Compared to Daya Bay Evaluation}

In this subsection we derive $\bar{\nu_e}$ disappearance, 
$\mathcal{P}(\bar{\nu}_e \rightarrow \bar{\nu}_e)$, defined as
\beq
\label{disappear}
  \mathcal{P}(\bar{\nu}_e \rightarrow \bar{\nu}_e)&=& 1-
 \mathcal{P}(\bar{\nu}_e \rightarrow \bar{\nu}_\mu)-
 \mathcal{P}(\bar{\nu}_e \rightarrow \bar{\nu}_\tau) \; ,
\eeq
using the S-matrix method (see previous subsection), and compare it
to the expression for $\mathcal{P}(\bar{\nu}_e \rightarrow \bar{\nu}_e)$
used by the Daya Bay, Double Chooz, and RENO, which is (see, e.g.,
Refs\cite{dc06,dc11})
\beq
\label{dc06P}
  \mathcal{P^{DB}}(\bar{\nu}_e \rightarrow \bar{\nu}_e)&\simeq & 1-
 4(s_{13}c_{13})^2 sin^2(\frac{\Delta L}{2})
\eeq
where $\Delta \equiv  \delta m_{13}^2/(2 E)$,  Eq(8), and $s_{13},c_{13}=
sin\theta_{13},cos\theta_{13}$. A third term with
a factor of $sin^2(\delta L/2)$\cite{dcnov11} was dropped\cite{dc06,dc11}
because $\delta m_{12}^2 \ll \delta m_{13}^2$ and $sin^2(\delta L/2)\ll
sin^2(\Delta L/2)$ for Daya Bay baseline L=1.9 km and energy E=4 Mev.

In the S-matrix method (see Ref\cite{jo04}) the probability of 
$\bar{\nu}_e$ oscillation to $\bar{\nu}_\mu$ and $\bar{\nu}_\tau$
is given by (see, e.g., Ref\cite{jo04})  
\beq
\label{barCP} 
 \mathcal{P^{SM}}(\bar{\nu}_e \rightarrow \bar{\nu}_\mu)&=&|\bar{S}_{21}|^2 
\nonumber \\
 \mathcal{P^{SM}}(\bar{\nu}_e \rightarrow \bar{\nu}_\tau)&=&|\bar{S}_{31}|^2 
\; .
\eeq
We take $\delta_{CP}=0$, since $|S_{12}|^2$ is essentially 
independent of $\delta_{CP}$, so $A=C$ (see Ref\cite{ahlo01} for definition
and proof). Therefore $|\bar{S}_{21}|^2$=$|S_{12}(V\rightarrow -V)|^2$,
and $|\bar{S}_{31}|^2$=$|S_{12}|^2(V\rightarrow -V,c_{23}\rightarrow s_{23},
 s_{23}\rightarrow -c_{23}$). Using $s_{23}^2=c_{23}^2= 1/2$ we find

\beq
\label{barSM}
 \mathcal{P^{SM}}(\bar{\nu}_e \rightarrow \bar{\nu}_e)&=& 1-
[(.46 \delta sin\bar{\omega} L/\bar{\omega})^2 +2(s_{13})^2(1- 
cos\bar{\bar{\Delta}} L)] 
\; ,
\eeq
with $\bar{\bar{\Delta}}=\Delta +(V-\delta)/2$ and 2 $\bar{\omega}=
 \sqrt{\delta^2 + V^2 +2 \delta V cos(2 \theta_{12})}$  

Fig. 2 for L=1.9km and Fig. 3 for L=10km are discussed below.
\clearpage
  
\begin{figure}[ht]
\begin{center}
\epsfig{file=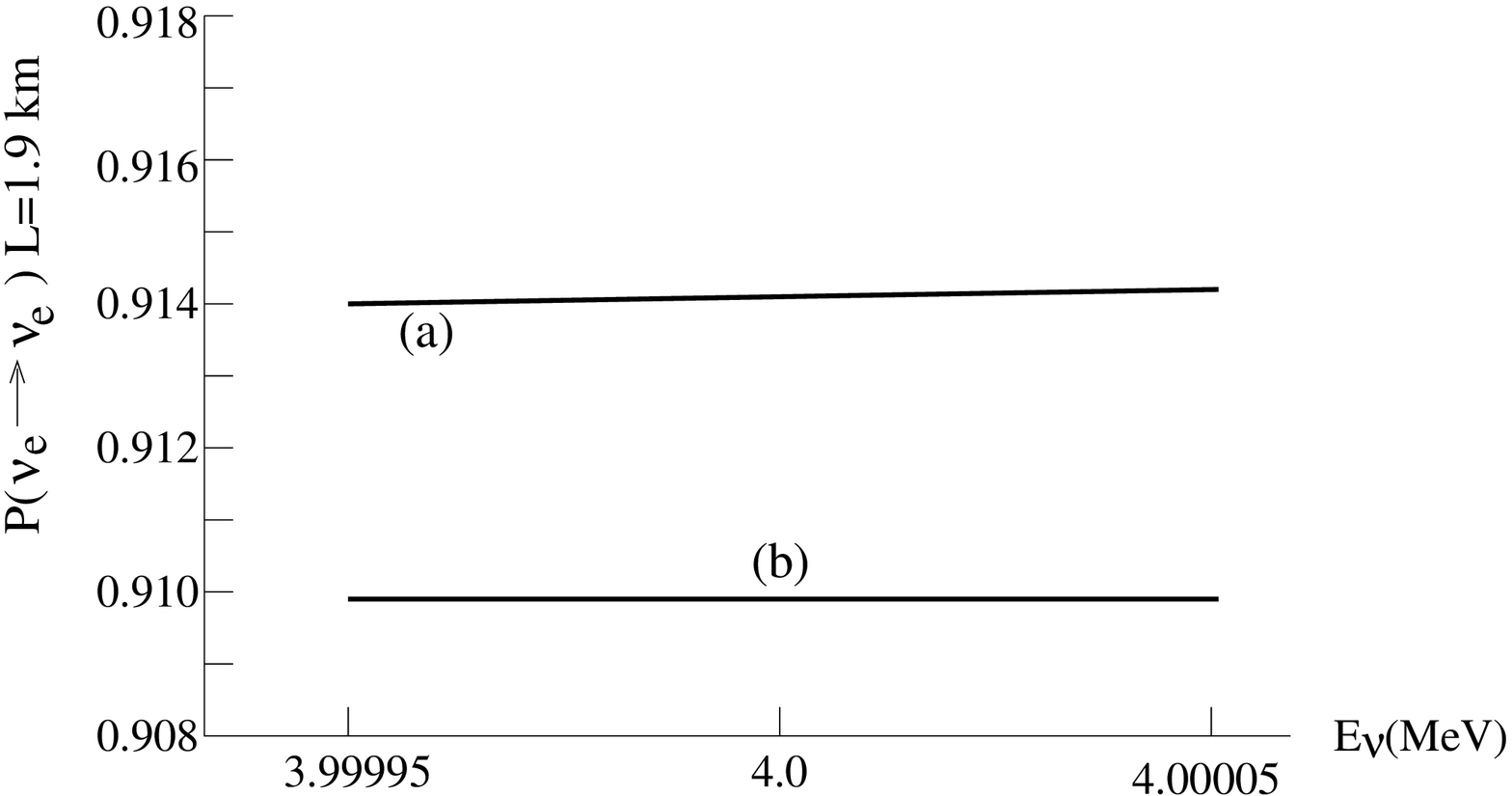,height=5cm,width=12cm}
\end{center}
\caption{$\mathcal{P}(\bar{\nu}_e \rightarrow \bar{\nu}_e)$. 
For $s_{13}=.15$ (a) $\mathcal{P^{DB}}$ and (b) $\mathcal{P^{SM}}$}
\end{figure}

  From Fig. 2 the ratios $R1, R2$, of $1-\mathcal{P}(\bar{\nu}_e \rightarrow 
\bar{\nu}_e)$ for $\mathcal{P}^{DB}$ (Eq(\ref{dc06P})) to $\mathcal{P}^{SM}$ 
(Eq(\ref{barSM}))
for $s_{13}=.15$, and for $s_{13}=.15$ for $\mathcal{P}^{DB}$ and $s_{13}=.147$
for $\mathcal{P}^{SM}$,  for E $\simeq$ 4.0MeV and L=1.9 km are
\beq
\label{ratio1}
   R1=\frac{1-\mathcal{P}^{DB}(s_{13}=.15)}{1-\mathcal{P}^{SM}(s_{13}=.15)}
&=& 1.04 \nonumber \\
   R2=\frac{1-\mathcal{P}^{DB}(s_{13}=.15)}{1-\mathcal{P}^{SM}(s_{13}=.147)}
&=& 1.00 \;,
\eeq
which demonstrates that using the S-Matrix formulation for L=1.9 km and 
E$\simeq$ 4.0 MeV one would extract $s_{13}=.147$ from the data for which
the older formalism finds $s_{13}=.15$. This is a 2\% correction. We use the
notation E$\simeq$ 4.0 MeV as there is uncertainty in the antineutrino
energy.
\vspace{1 cm}

\begin{figure}[ht]
\begin{center}
\epsfig{file=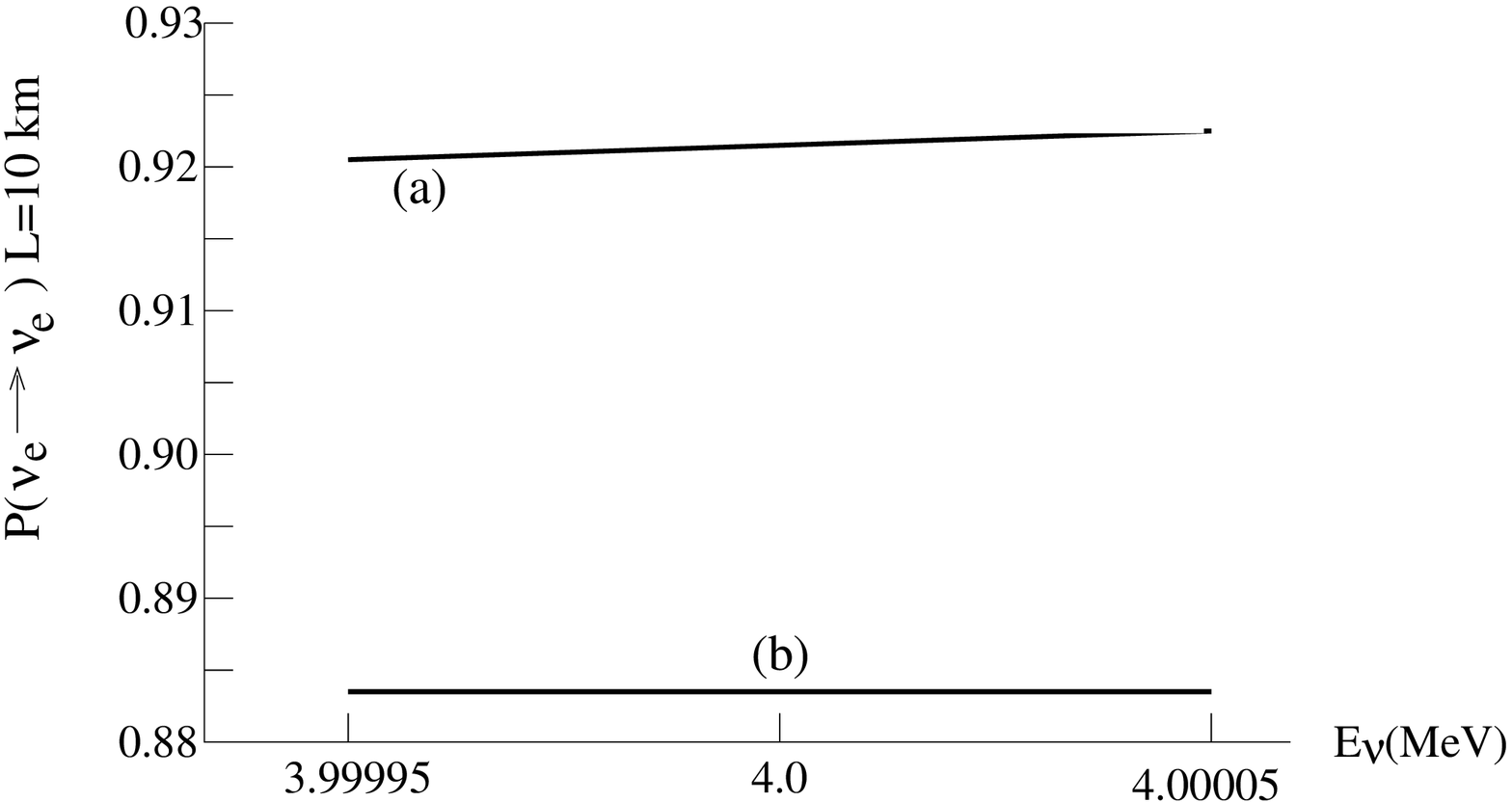,height=5cm,width=12cm}
\end{center}
\caption{$\mathcal{P}(\bar{\nu}_e \rightarrow \bar{\nu}_e)$. 
For $s_{13}=.15$ (a) $\mathcal{P^{DB}}$ and (b) $\mathcal{P^{SM}}$}
\end{figure}

  Fig. 3 is the same as Fig. 2, except the baseline is L=10 km, as future
might use a longer baseline for a larger effect given $s_{13}$. 
For E $\simeq$ 4.0MeV and L=10 km the ratios are
\beq
\label{ratio1}
   R1=\frac{1-\mathcal{P}^{DB}(s_{13}=.15)}{1-\mathcal{P}^{SM}(s_{13}=.15)}
&=& 1.47 \nonumber \\
   R2=\frac{1-\mathcal{P}^{DB}(s_{13}=.15)}{1-\mathcal{P}^{SM}(s_{13}=.095)}
&=& 1.00 \;.
\eeq
Thus using the S-Matrix formulation for L=10 km and 
E$\simeq$ 4.0 MeV one would extract $s_{13}=.095$ from the data for which
the older formalism finds $s_{13}=.15$. This is a 35\% correction.

We have carried out similar calculations for the 
T2K project, with E=0.6 GeV, L=295 km\cite{T2K11}. With both a larger L and 
larger E than Daya Bay,  we find a correction of 2.4\%.

   It is also important to note that our SM method
gives $\mathcal{P}(\bar{\nu}_e\rightarrow \bar{\nu}_e)\neq 1.0$ even for
$s_{13}$=0. 
\section{CP Violation $\Delta \mathcal{P}^{CP}_{\mu e}$}

   In this section we shall extend the derivation of the transition
probability $\mathcal{P}(\nu_\mu \rightarrow \nu_e)$ of the previous
section to derive the CPV probability
\beq
\label{CPV}
  \Delta\mathcal{P}^{CP}_{\mu e} &=& \mathcal{P}(\nu_\mu \rightarrow \nu_e)
-\mathcal{P}(\bar{\nu}_\mu \rightarrow \bar{\nu}_e) \nonumber \\
         &=&  |S_{12}|^2- |\bar{S}_{12}|^2 \,
\eeq
with $S_{12}$ defined in Eq(8) and
\beq
\label{S12bar}
  \bar{S}_{12} &=&  c_{23} \bar{\beta} -is_{23} a e^{i\delta_{CP}} \bar{A}\; ,
\eeq
with $\bar{\beta}= \beta (V \rightarrow -V)$ and $\bar{A}= A(V \rightarrow -V)$.
For example (see Eqs(16,18)) $2\bar{\omega}= \sqrt{\delta^2 + V^2 +2 \delta V 
cos(2 \theta_{12})}$ and $\bar{\bar{\Delta}}=\Delta+(V-\delta)/2$. Using
conservation of probabiltiy\cite{jo04}, $|A|^2=|\bar{A}|^2$.
\beq
\label{DCPV}
  \Delta\mathcal{P}^{CP}_{\mu e} &=& c_{23}^2(|\beta|^2-|\bar{\beta}|^2)
-2 c_{23} s_{23} ai (Im[\beta e^{i\delta_{CP}} A^*]
-Im[\bar{\beta}e^{i\delta_{CP}} \bar{A}^*]) \; .
\eeq

\newpage

From Eq(\ref{DCPV}), the definitions in the previous section, defining
$s \equiv sin(\omega L)$, $c \equiv cos(\omega L)$ one finds
\beq
\label{DCPVf}
  \Delta\mathcal{P}^{CP}_{\mu e} &=& c_{23}^2 s_{12}^2 c_{12}^2 \delta^2
(\frac{s^2}{\omega^2}-\frac{\bar{s}^2}{\bar{\omega}^2}) +2 c_{23}s_{23}
s_{12}c_{12}s_{13}\delta (\Delta-\delta s_{12}^2)  \\
  &&(sin\theta_{CP}(\frac{s}{\omega}(c-cos\bar{\Delta}L)
\frac{\bar{\Delta}-\omega cos 2\theta}
{\bar{\Delta}^2-\omega^2}+\frac{\bar{s}}{\bar{\omega}}(\bar{c}-
cos\bar{\bar{\Delta}}L)
 \frac{\bar{\bar{\Delta}}-\bar{\omega} cos 2\bar{\theta}}
{\bar{\bar{\Delta}}^2-\bar{\omega}^2})) \nonumber \\
&&- cos\theta_{CP}(\frac{s}{\omega}\frac{sin\bar{\Delta}L(\bar{\Delta}
-\omega cos2\theta)+sin \omega L(\omega+\bar{\Delta}cos 2\theta)}
{\bar{\Delta}^2-\omega^2}\\
&&-\frac{\bar{s}}{\bar{\omega}}\frac{sin\bar{\bar{\Delta}}L(\bar{\bar{\Delta}}
-\bar{\omega} cos2\theta)+sin\bar{\omega} L(\bar{\omega}
+\bar{\bar{\Delta}}cos 2\theta)}{\bar{\bar{\Delta}}^2-\bar{\omega}^2})
\; .
\eeq

 The results for $\Delta\mathcal{P}^{CP}_{\mu e}$ for $s_{13}$=.19 are shown 
in Fig.4. Note that $\Delta\mathcal{P}^{CP}_{\mu e}$ depends strongly on
$\delta_{CP}$, which could lead to a measurement of this parameter. The
large value of $\Delta\mathcal{P}^{CP}_{\mu e}$ for CHOOZ is promising
for future experiments. $\Delta\mathcal{P}^{CP}_{\mu e}$ is so small (from
about $10^{-10}$ to $10^{-18}$)for MiniBooNE, we do not show the results. 
Similar results for $\Delta\mathcal{P}^{CP}_{\mu e}$ for $s_{13}$=.095 are shown 
in Fig.5.
\newpage

\begin{figure}[ht]
\begin{center}
\epsfig{file=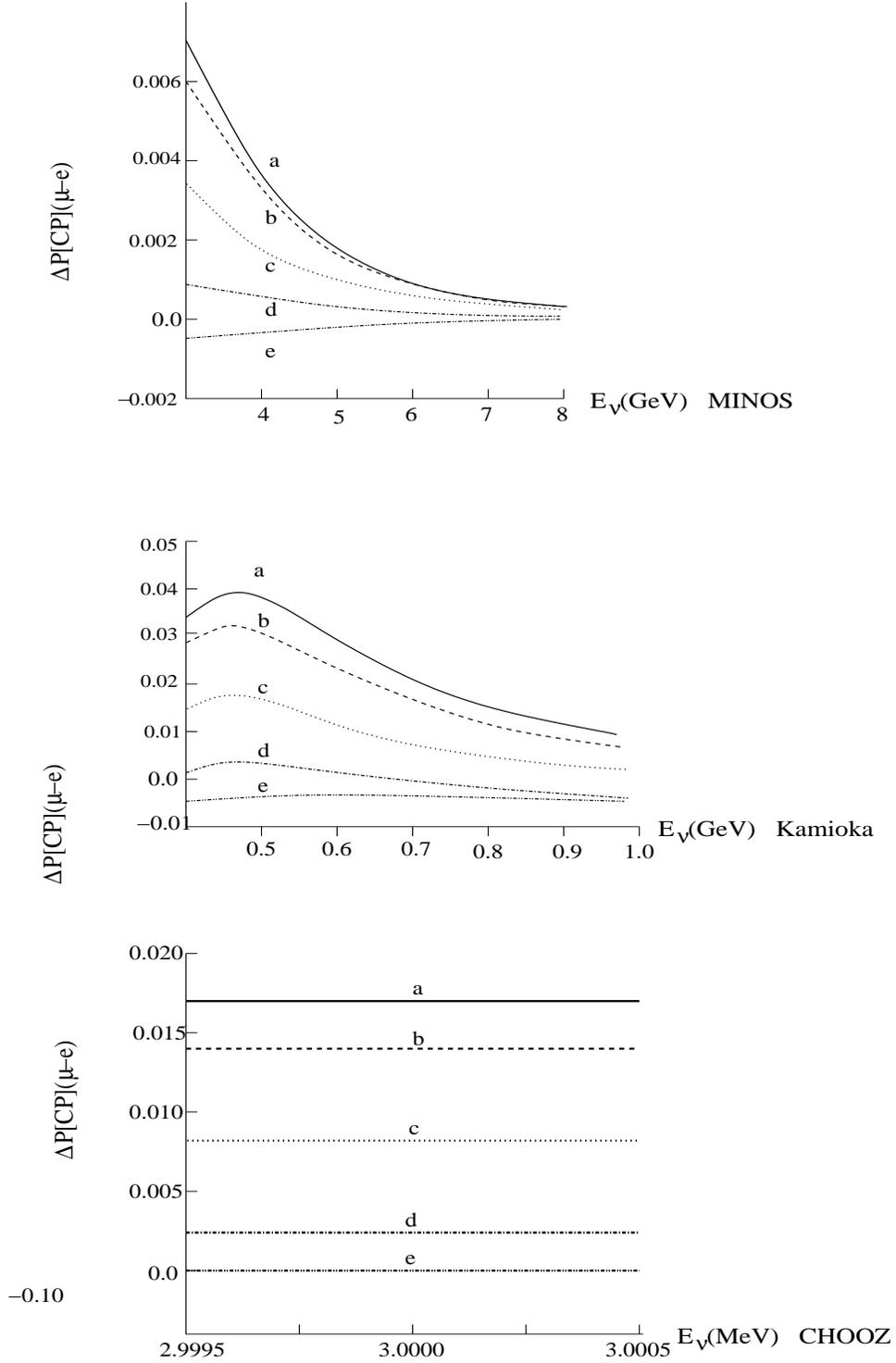,height=19.5cm,width=13cm}
\end{center}
\caption{The ordinate is $\Delta \mathcal{P}(\nu_\mu \rightarrow\nu_e)$ for 
MINOS(L=735 km), JHF-Kamioka(L=295 km), and CHOOZ(L=1 km).
s13=.19, and a, b, c, d, e for $\delta_{CP}$=$\pi/2$, $\pi/4$,
0.0, $-\pi/4$, $-\pi/2$}
\end{figure}

\newpage

\begin{figure}[ht]
\begin{center}
\epsfig{file=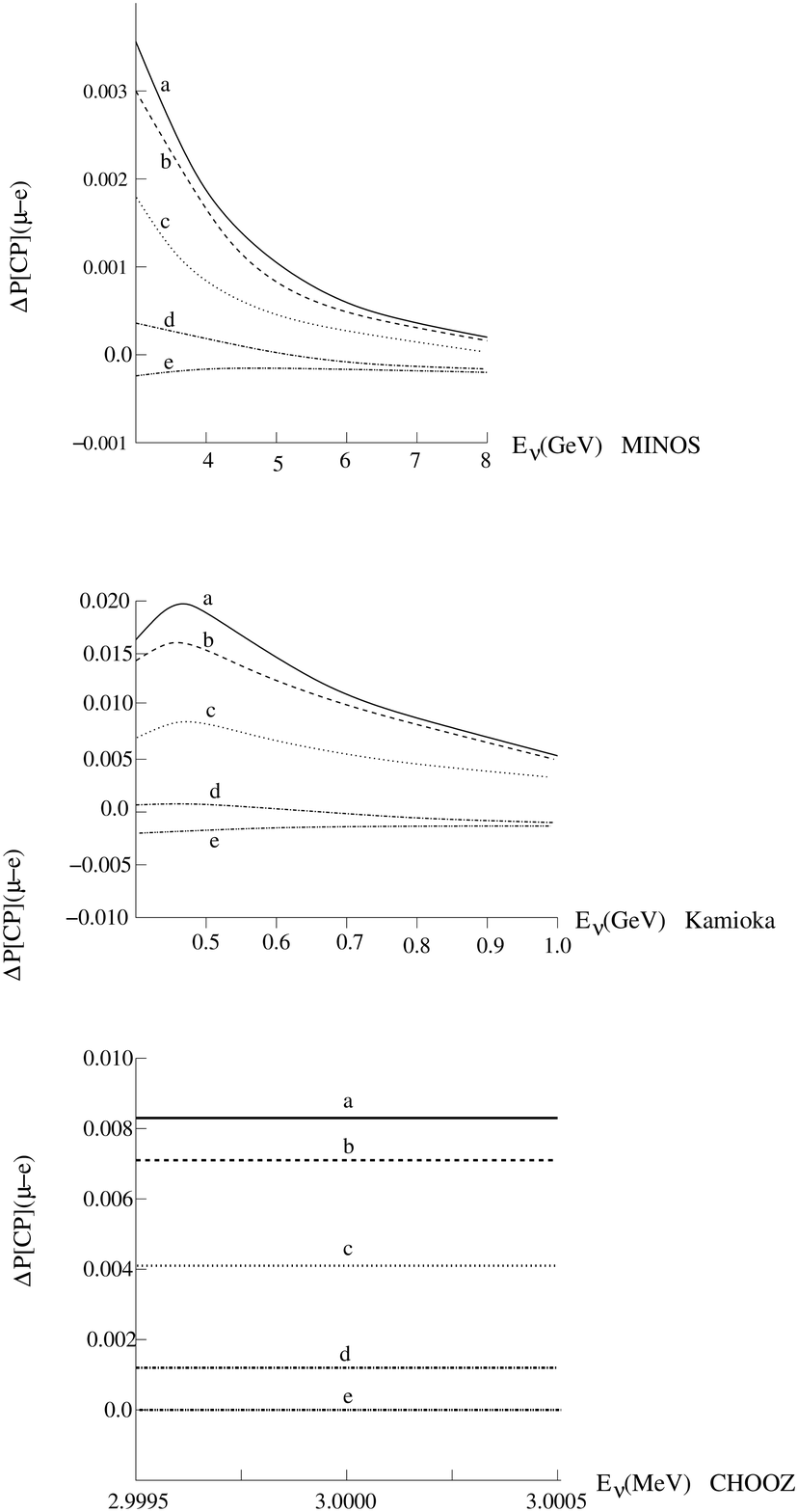,height=19.5cm,width=13cm}
\end{center}
\caption{The ordinate is $\Delta \mathcal{P}(\nu_\mu \rightarrow\nu_e)$ for 
MINOS(L=735 km), JHF-Kamioka(L=295 km), and CHOOZ(L=1 km).
s13=.095, and a, b, c, d, e for $\delta_{CP}$=$\pi/2$, $\pi/4$,
0.0, $-\pi/4$, $-\pi/2$}
\end{figure}

\clearpage
\section{Conclusions} 

We have estimated CP violation for a variety of experimental neutrino beam
facilities, for values of the parameter $s_{13}$ =0.19 and .095, and for 
$\delta_{CP}$ from 90 to -90 degrees, since its value is not known. As our 
results show, the probability $\mathcal{P}(\nu_\mu \rightarrow\nu_e)$ is 
strongly dependent on $s_{13}$ and is essentially independent of $\delta_{CP}$ 
(see Fig. 1), and therefore the measurement of $\mathcal{P}(\nu_\mu 
\rightarrow\nu_e)$ should determine the value of the $s_{13}$ parameter, 
as has been known for many years. 

Our new results for $\bar{\nu}_e$ disappearance, as is being measured the 
Daya Bay, Double Chooz and RENO projects, however, make use of a different 
theoretical formulation than that used by these projects to extract $s_{13}$ 
from the data. We have shown that the recent result from the Daya Bay 
collaboration\cite{DB3-7-12} with E$\simeq$4 MeV and L=1.9 km, from which it 
was estimated that $s_{13}\simeq .15$, by our analysis is $s_{13}\simeq .147$, 
a 2\% correction. This is small, but the goal of these projects is 1\% 
accuracy for $s_{13}$. For a baseline of L=10 km, with E$\simeq$ 4 MeV, we 
find $s_{13}\simeq .097$ using the S-Matrix method, rather than .15, a 35\%
correction. Also, our SM method gives 
$\mathcal{P}(\bar{\nu}_e\rightarrow \bar{\nu}_e)\neq 1.0$ even for
$s_{13}$=0.

The CP violation probability (CPV), $\Delta\mathcal{P}^{CP}_{\mu e}$, 
is strongly dependent on both of these important parameters.
After the Double Chooz/Daya Bay/RENO determination of $s_{13}$, both the
JHF-Kamioka and Double Chooz projects might be able to determine the value 
of $\delta_{CP}$, since for most of the values of $\delta_{CP}$ these 
projects would have nearly a 1\% CPV, as shown in Figs. 4 and 5.
No experiments are possible now to test CPV via neutrino 
oscillations, since beams of both neutrino and antineutrino with the same
flavor would be needed. However, in the future such beams might be available.
Our results should help in planning such future experiments.

\vspace{3mm}

\Large{{\bf Acknowledgements}}\\
\normalsize
This work was supported in part by the NSF grant PHY-00070888, in part 
by the DOE contracts W-7405-ENG-36 and DE-FG02-97ER41014, and in part 
by a grant from the Pittsburgh Foundation. We thank Prof Ma Wei-xing,
IHEP Beijing, for information about the Daya Bay project.

\end{document}